\documentclass{elsart}
\usepackage{amssymb,latexsym}

\begin{document}

\newtheorem{theorem}{Theorem} 
\renewcommand{\theequation}{\arabic{section}.\arabic{equation}}

\begin{frontmatter}

\title{Maximal Entanglement of Nonorthogonal States: Classification}

\author[fu]{Hongchen Fu\corauthref{cor}} 
\author[wang]{Xiaoguang Wang}
\author[fu]{Allan I Solomon}

\address[fu]{Quantum Processes Group, The Open University, 
Milton Keynes, MK7 6AA, United Kingdom}
\corauth[cor]{Corresponding author.}
\ead{h.fu@open.ac.uk, a.i.solomon@open.ac.uk, xgwang@isiosf.isi.it}

\address[wang]{Quantum Information Group, 
             Institute for Scientific Interchange (ISI) Foundation, 
             Viale Settimio Severo 65, I-10133 Torino, Italy }

\begin{abstract}
A necessary and sufficient condition for the maximal entanglement 
of  bipartite  nonorthogonal pure states is found.  The condition
is applied  to the maximal entanglement of coherent states. 
Some new classes of maximally entangled coherent states  
 are explicited constructed; their limits give rise 
 to maximally entangled Bell-like states.
\end{abstract}

\end{frontmatter}

\section{Introduction}

Quantum entanglement plays an important role 
in such areas of quantum information processing  as quantum
teleportation \cite{ben1}, superdense coding \cite{ben2},
quantum key distribution \cite{eke} and telecloning 
\cite{mur}. Recently, entangled nonorthogonal states 
have attracted much attention in quantum cryptography \cite{fuch}. 
Bosonic entangled coherent states \cite{sand,enk,hir1,hir2,jeo} 
and $su(2)$ and 
$su(1,1)$ entangled coherent states \cite{wang1} are typical 
examples of entangled nonorthogonal states. For general 
bipartite nonorthogonal states some conditions have been  
found for  maximal entanglement \cite{wang2}. 

In this letter we address the problem of  finding  necessary and 
sufficient conditions for the  maximal entanglement of  bipartite 
nonorthogonal pure states \cite{pere,mann}
\begin{equation} \label{nstate}
|\psi\rangle = \mu |\alpha\rangle\otimes|\beta\rangle +
\nu |\gamma\rangle\otimes|\delta\rangle, 
\end{equation}
where $\mu, \nu$ are two complex numbers, $|\alpha\rangle$
and $|\gamma\rangle$ are linearly independent normalized 
states of the systems 1, and $|\beta\rangle$ and $|\delta\rangle$ 
are linearly independent normalized states of the system 2. Here
we are interested in the nonorthogonal case in which the 
overlaps $\langle\alpha|\gamma\rangle$ and
 $\langle\beta|\delta\rangle$ are non-vanishing.
 Note that the normalization constant is not 
 $|\mu|^2+|\nu|^2$, but 
 \begin{equation}
 \left[ |\mu^2|+|\nu|^2 + 
 \mu\nu^* \langle \gamma|\alpha\rangle\langle
 \delta|\beta\rangle
 +\mu^*\nu \langle\alpha|\gamma\rangle
 \langle\beta|\delta\rangle\right]^{-1/2}.
 \end{equation}
 We first give a classification of maximal entanglement 
 for the state (\ref{nstate}). A necessary and sufficient 
 condition for maximal entanglement is found.  We then
 apply the condition to 
 coherent states and find a simple  condition for maximally entangled coherent 
 states in terms of the coherence parameters. Some new classes of maximally
 entangled coherent states, with relative phase different from the 
 previously-noted $\pi$, are explicited constructed.  In the limit, these states give rise  to maximally entangled Bell-like states.

\section{Maximal entanglement: classification}
\setcounter{equation}{0}

The entanglement of a quantum system can be measured by the
{\em entanglement of formation}, or simply {\em entanglement},
which is defined as the entropy of either one of the two subsystems 1
or 2 \cite{5320}.
It has been  pointed out that the entanglement of a two-qubit state 
$|\Psi\rangle$ can be expressed as a 
function of the {\em concurrence} \cite{conc}
\begin{equation}
C\equiv |\langle \Psi|\sigma_y \otimes \sigma_y|\Psi^*\rangle|,
\end{equation}
where $\sigma_y$ is the {\em spin-flip} operator and 
$|\Psi^*\rangle$ is the complex conjugate of $|\Psi\rangle$.
Concurrence itself can also be regarded as a  
measure of entanglement which ranges from 0 to 1 
\cite{conc}. One may readily obtain the concurrence for the state (\ref{nstate}) by
introducing an orthogonal normalized basis in the subspace
spanned by $|\alpha\rangle$ and $|\gamma\rangle$ and 
by $|\beta\rangle$ and $|\delta\rangle$; it has the value 
\cite{wang2}
\begin{equation}
C=\frac{2|\mu\nu|\sqrt{(1-|\langle\alpha|\gamma\rangle|^2)(
1-|\langle\beta|\delta\rangle|^2)}}{|\mu|^2 +|\nu|^2+\mu\nu^*
\langle\gamma|\alpha\rangle\langle\delta|\beta\rangle+
\mu^*\nu \langle\alpha|\gamma\rangle\langle\beta|\delta\rangle}.
\end{equation}
The state (\ref{nstate}) is referred to as a {\em Maximally Entangled 
State} (MES) when $ C=1$.

We now  give a necessary and sufficient condition 
for  maximal entanglement ($C=1$).
Let $\mu=k\nu e^{i\theta}$, where $k$
and $\theta$ be real parameters with $k>0$. Noting 
that $|\langle\alpha|\gamma\rangle|\leq 1$ and 
$|\langle \beta | \delta \rangle|\leq 1$, we write  
\begin{equation}
\langle\alpha|\gamma\rangle=\sin a \, e^{i\theta_1}, \qquad
\langle \beta | \delta \rangle =\sin b \, e^{i\theta_2}, 
\end{equation}
where $a, b$ and $\theta_1, \theta_2$ are all real parameters
and $0 \leq a,b\leq \pi/2$. Then we may rewrite  the condition   $C=1$ as
\begin{equation} \label{condition2}
k' = 2\cos a \cos b - 2\sin a \sin b \cos(\theta-\theta_1-\theta_2),
\end{equation}
where $k'\equiv (k^2+1)/k$.
We consider two different cases.


\noindent {\bf Case 1:} $-1\leq \cos(\theta-\theta_1-\theta_2) \leq 0$.
In this case we have
\begin{equation} \label{kprime1}
k' \leq 2\cos a \cos b + 2\sin a \sin b = 2\cos(a-b)\leq 2,
\end{equation}
namely $(k-1)^2\leq 0$, or $k=1$, $k'=2$. Inserting $k'=2$ into
relation (\ref{kprime1}) we also have $\cos(a-b)=1$, namely
$a=b+2m\pi$ ($m$ an integer). Then from Eq.(\ref{condition2}) 
we have
\begin{equation}
\cos(\theta-\theta_1-\theta_2)=-1,
\end{equation}
namely $\theta-\theta_1-\theta_2=\pi$. So in this case the
MES condition is obtained as
\begin{equation} \label{mescon}
\mu=\nu e^{i\theta} , \qquad
\langle\alpha|\gamma\rangle= 
-\langle \beta | \delta \rangle^* e^{i\theta}. 
\end{equation}

\noindent {\bf Case 2:} $0\leq \cos(\theta-\theta_1-\theta_2) \leq 1$.
In this case the second term in (\ref{condition2}) is always 
non-negative and thus
\begin{equation} \label{kprime2}
k'\leq 2 \cos a \cos b \leq 2,
\end{equation}
which leads to $k=1$ and $k'=2$ as in case 1. Then the relation
(\ref{kprime2}) is valid only when 
\begin{equation}
\cos a =\cos b =1 \mbox{ or } \sin a =\sin b= 0.
\end{equation}
So the MES condition in this case is
\begin{equation}
|\mu| = |\nu|, \quad
\langle\alpha|\gamma\rangle=\langle \beta | \delta \rangle=0,
\end{equation}
which is clearly the orthogonal case.

In summary we obtain the following theorem

\begin{theorem}
The states (\ref{nstate}) are MES if and only if one of the 
following conditions is satisfied
\begin{enumerate}
\item  $\mu=\nu e^{i\theta}$ and   
$\langle\alpha|\gamma\rangle = -\langle \beta | \delta \rangle^* e^{i\theta}$
($\theta$ is a real parameter) for the nonorthogonal case;

\item $|\mu|=|\nu|$ for the orthogonal case
$\langle\alpha|\gamma\rangle = \langle \beta | \delta \rangle=0$.
\end{enumerate}
\end{theorem}

The necessity can be verified directly.

Before closing this section we remark that the state (\ref{nstate})
is disentangled ($C=0$) if and only if one of the following is true: 
(1) $\mu=0$;
(2) $\nu=0$; (3) $|\alpha\rangle = \pm |\gamma\rangle$;
(4) $|\beta\rangle = \pm |\delta\rangle$.  

\section{Maximally entangled states}
\setcounter{equation}{0}

Now we turn to the explicit construction of MES. It is easy to see 
that  condition (\ref{mescon}) is satisfied when
\begin{equation}
|\gamma\rangle = e^{i \vartheta}|\beta\rangle, \qquad
|\delta\rangle = e^{i(\theta+\pi-\vartheta)} |\alpha\rangle,
\end{equation}
where $\vartheta$ is an arbitrary phase. In this case, the 
normalized MES is obtained as
\begin{equation}
|\psi\rangle = 
\frac{\mu}{|\mu|\sqrt{2(1-|\langle\alpha|\beta\rangle|^2 )}}
(|\alpha\rangle\otimes|\beta\rangle - 
|\beta\rangle\otimes|\alpha\rangle),
\end{equation}
which is just the {\em antisymmetric MES} given in paper
\cite{wang2}. Note that the phase $e^{i\theta}$ does not
enter the anti-symmetric state. It is natural to ask if there 
exist  MES other than the antisymmetric MES. 
The answer is positive.

Let us consider  entangled coherent states; namely, all 
four states in (\ref{nstate}) are coherent states
\begin{equation}
|\alpha\rangle = e^{-|\alpha|^2/2} \sum_{n=0}^\infty
\frac{\alpha^n}{\sqrt{n!}}|n\rangle,
\end{equation}
where $\alpha$ is an arbitrary complex number.
The overlap between two coherent states 
$|\alpha\rangle$ and $|\gamma\rangle$ is
\begin{equation}
\langle \alpha |\gamma\rangle = \exp\left[
-\frac{1}{2} \left(|\alpha|^2 + |\gamma|^2 - 
2\alpha^*\gamma \right)\right].
\end{equation}
Then the MES condition (\ref{mescon})  simplifies to
\begin{equation}  \label{cohcon}
|\alpha^2 + |\gamma|^2 - 2 \alpha^* \gamma =
|\beta|^2 + |\delta|^2 -2 \beta \delta^* - 2 i (\theta+\pi),
\end{equation}
from which we conclude that
\begin{theorem}
Coherent states are maximally entangled 
if and only if $\mu=\nu \exp(i \theta) $ and
both sides of Eq.(\ref{cohcon}) have the 
same {\em real} part
\begin{equation}
|\alpha^2 + |\gamma|^2 - 2\rm{Re}(\alpha^* \gamma) = 
|\beta|^2 + |\delta|^2 -2\rm{Re}(\beta \delta^*). 
\end{equation}
The difference of their imaginary parts gives rise to a 
relative phase $\theta$
\begin{equation}
\theta = \rm{Im}(\alpha^* \gamma) - 
\rm{Im}(\beta \delta^*) - \pi. 
\end{equation}
\end{theorem}

As an example we can choose $\alpha$ and $\delta$ such that they 
have the same phase, as do  $\beta$ and $\delta$. In this 
case $\alpha^*\gamma = |\alpha||\gamma|$,
$\beta\delta^* = |\beta||\delta|$ and $\theta=-\pi$. Then the MES
condition is obtained as
\begin{equation}
|\alpha|-|\gamma|=\pm (|\beta|-|\delta|) = \lambda',
\end{equation}
where $\lambda'$ is a {\em real} parameter, and the 
corresponding MES are
\begin{equation}
      |\alpha\rangle \otimes |\beta\rangle -
      \left|(1-\lambda'/|\alpha|)\alpha \right\rangle \otimes
      \left|(1\mp \lambda'/|\beta|)\beta\right\rangle.
\end{equation}  
If we further choose $\beta=-\alpha$ and $\lambda'=2|\alpha|$,
we obtain the well-known MES
\begin{equation} \label{wellknown}
      |\alpha\rangle\otimes|-\alpha\rangle-
      |-\alpha\rangle\otimes|\alpha\rangle
\end{equation}           
and a new MES
\begin{equation} \label{statewith3}
      |\alpha\rangle\otimes|-\alpha\rangle-
      |-\alpha\rangle\otimes|-3\alpha\rangle.
\end{equation}    
Both states (\ref{wellknown}) and (\ref{statewith3}) have the 
same normalization constant 
$[2(1-e^{-4|\alpha|^2})]^{-\frac{1}{2}}$.

We now give an example in which the relative phase 
is not $\pi$. Suppose that $\alpha$ and $\gamma$ have a phase difference of
$\pi/2$, as therefore   do $\beta$ and $\gamma$,
namely;
\begin{equation}
\frac{\gamma}{|\gamma|}=\frac{\alpha}{|\alpha|}\, e^{i\pi/2}, \qquad
\frac{\delta}{|\delta|}=\frac{\beta}{|\beta|} \, e^{\pm i\pi/2}, 
\end{equation}
Then $\alpha^*\gamma$ and $ \beta \delta^*$ are pure imaginary
\begin{equation}
\alpha^*\gamma = i|\alpha\gamma|, \qquad
\beta \delta^* = \mp i|\beta\delta|.
\end{equation}
So the MES conditions in this case are
\begin{eqnarray}
&& |\alpha|^2 + |\gamma|^2 = |\beta|^2 + |\delta|^2 , \\
&& \theta=|\alpha \gamma|\pm |\beta\delta| -\pi ,
\end{eqnarray}
and the MES is obtained as
\begin{equation}
|\alpha\rangle\otimes|\beta\rangle - 
e^{-i(|\alpha\gamma|\pm |\beta\delta|)}\,
\left| i |\gamma| \alpha / |\alpha| \right\rangle\otimes
\left|\pm i |\delta| \beta /|\beta|\right\rangle.
\end{equation}
In the case $|\alpha|=|\beta|=|\gamma|=|\delta|$, the
MES states  further simplify to
\begin{equation}
|\alpha\rangle\otimes|\beta\rangle - 
e^{-i(|\alpha|^2 \pm |\alpha|^2)}\,
\left| i \alpha \right\rangle\otimes
\left|\pm i \beta \right\rangle.
\end{equation}
In particular, when $\alpha= \pm \beta$, we  obtain 
some new types of MES
\begin{eqnarray}
&& |\alpha\rangle\otimes|-\alpha\rangle - 
| i \alpha \rangle\otimes |  i \alpha \rangle,
\label{4states1}\\
&& |\alpha\rangle\otimes|-\alpha\rangle - 
e^{- i 2|\alpha|^2}\,
| i \alpha \rangle\otimes | -i \alpha \rangle,
\label{4states2}\\
&& |\alpha\rangle\otimes| \alpha\rangle - 
| i \alpha \rangle\otimes | -i \alpha \rangle,
\label{4states3}\\
&& |\alpha\rangle\otimes|\alpha\rangle - 
e^{-i 2|\alpha|^2}\,
| i \alpha \rangle\otimes |  i \alpha \rangle.
\label{4states4}
\end{eqnarray} 
These all have the same normalization constant
\begin{equation}
\frac{1}{\sqrt{2(1-e^{-2|\alpha|^2})}}.
\end{equation}
It is interesting that the last two states (\ref{4states3}) 
and (\ref{4states4})  can be obtained from the first two 
states (\ref{4states1}) and (\ref{4states2}), 
respectively, by the local transformation
$1\otimes (-1)^{a^{\dagger}_2 a_2}$, where $a^\dagger_2$ and
$a_2$ are creation and annihilation operators of the
second harmonic oscillator system respectively.

Note that the property of maximal entanglement of the 
above states is independent of the value of $\alpha$. 
Let us consider the state (\ref{4states1}).  We expand it 
in  Fock space as
\begin{equation}
\frac{e^{-|\alpha|^2}}{\sqrt{2(1-e^{-2|\alpha|^2})}}
\sum_{m,n=0}^\infty 
\frac{\alpha^{m+n}}{\sqrt{m!\, n!}}\left[ (-1)^n -i^{m+n}
\right]|m\rangle\otimes |n\rangle,
\end{equation}
in which only terms with $m+n=1$ survive in the limit
$|\alpha|\to 0$ (there is {\em no} term $|0\rangle\otimes|0\rangle$).
The limiting state is readily  obtained as
\begin{equation}
\frac{1}{\sqrt{2}}\left( e^{i\pi/4} |0\rangle\otimes |1\rangle -
e^{-i\pi/4} |1\rangle\otimes |0\rangle \right),
\end{equation}
which is clearly a Bell-like MES. We may  similarly show that
the states (\ref{4states2} - \ref{4states4}) degenerate to
the following orthogonal Bell-like MES in the limit $|\alpha|\to 0$
\begin{eqnarray}
&& 
\frac{1}{\sqrt{2}}\left( e^{i\pi/4} |0\rangle\otimes |1\rangle +
e^{-i\pi/4} |1\rangle\otimes |0\rangle \right),\\
&& \frac{1}{\sqrt{2}}\left( |0\rangle\otimes |1\rangle -
|1\rangle\otimes |0\rangle \right), \\
&&  \frac{1}{\sqrt{2}}\left( |0\rangle\otimes |1\rangle +
|1\rangle\otimes |0\rangle \right), \label{323}
\end{eqnarray}
respectively, and the states Eq.(\ref{wellknown})
and Eq.(\ref{statewith3}) degenerate to (\ref{323}).

\section{Conclusion}

In this letter we  gave  necessary and sufficient conditions
for the maximal entanglement of bipartite nonorthogonal pure 
states (\ref{nstate}). We then applied  these conditions to  
entangled coherent states and explicitly constructed some new 
types of maximally entangled coherent states. Apart from the 
antisymmetric example (\ref{wellknown}), these maximally 
entangled coherent states are novel in that they have
 relative phases other than $\pi$.  In the limit when the coherence parameter tends to zero,  these  maximally 
 entangled coherent states give rise to maximally entangled Bell-like states.   We intend to generalize this formalism 
 to the case of mixed states and consider possible  applications 
 in the area of quantum information.

\section*{Acknowledgement}

H.\,Fu is supported in part by 
the National Natural Science Foundation of China
(19875008) and X.\,Wang is supported by the European 
Project Q-ACTA.
A.\,I.\,Solomon acknowledges the hospitality of the 
Laboratoire de Physique Th\'{e}orique des Liquides, 
Paris University VI.

\end{document}